\documentstyle[preprint,eqsecnum,aps,prb]{revtex}

\begin{document}
\title{Effect of ferroelectric layers on the magnetocapacitance properties of
superlattices-based oxide multiferroics }
\author{M.P.\ Singh$^1$, W. Prellier\thanks{%
prellier@ensicaen.fr}$^{,1}$, L.\ Mechin$^2$ and B.\ Raveau$^1$.}
\address{$^1$Laboratoire CRISMAT, CNRS\ UMR 6508, ENSICAEN,\\
6 Bd du Mar\'{e}chal Juin, F-14050 Caen Cedex, FRANCE.\\
$^2$Laboratoire GREYC, CNRS\ UMR 6072, ENSICAEN and University of Caen, 6 Bd%
\\
du Mar\'{e}chal Juin, F-14050 Caen Cedex, FRANCE.}
\date{\today}
\maketitle

\begin{abstract}
A series of superlattices composed of ferromagnetic La$_{0.7}$Ca$_{0.3}$MnO$%
_3$ (LCMO) and ferroelectric/paraelectric Ba$_{1-x}$Sr$_x$TiO$_3$ (0$\leq $x$%
\leq $1) were deposited on SrTiO$_3$ substrates using the pulsed laser
deposition. Films of epitaxial nature comprised of spherical mounds having
uniform size are obtained. Magnetotransport properties of the films reveal a
ferromagnetic Curie temperature in the range of 145-158 K and negative
magnetoresistance as high as 30\%, depending on the type of ferroelectric
layers employed for their growth ({\it i.e.} '{\it x'} value).
Ferroelectricity at temperatures ranging from 55 K to 105 K is also
observed, depending on the barium content. More importantly, the
multiferroic nature of the film is determined by the appearance of negative
magnetocapacitance, which was found to be maximum around the ferroelectric
transition temperature (3\% per {\it tesla}). These results are understood
based on the role of the ferroelectric/paraelectric layers and strains in
inducing the multiferroism.
\end{abstract}

\newpage

After a long gap, interest has been renewed in multiferroic materials, where
ferromagnetism and ferroelectricity coexist \cite{Schmid} and possess
multiferroic coupling, {\it i.e.,} magnetic domains can be switched by
applied electric field and likewise electric domains are switched by applied
magnetic field. As a consequence, the latter are potential candidates for
various novel devices, {\it e.g., }multiple state memory elements and
electric field controlled ferromagnetic devices. There are two open basic
issues in this field of research: first, what is the origin of multiferroism
and second, how to enhance them for suitable applications? Since very few
multiferroic materials exist in nature or are synthesized in laboratory\cite
{wang,Lorenz,kimura,zheng}, the scarcity of such materials is a big
stumbling block to address the aforementioned issues\cite{WP}. It is
therefore timing and warranting to design novel multiferroics and
characterize them for their essential properties.

To design a new (artificial) multiferroic, usually three distinct approaches
can be adopted, which can be described as follows. The first one deals with
the doping of suitable elements {\it e.g.}, ferromagnetic element(s) in
ferroelectric host and vice versa. The second is concerning the designing of
composites with the ferroelectric and magnetic hosts. The third consists in
designing superlattices composed of either bi-components ({\it e.g.,} one is
ferroelectric and other is ferromagnetic layer, or a nanocomposites\cite
{zheng}) or of multi-components ({\it i.e.,} with three or more distinct
compound layers) where one will observe the breaking symmetry by which it
may be possible to induce ferroelectricity (FE) and/or ferromagnetism (FM).
In building the superlattices for multiferroics, the total structure should
be an insulator, but FM materials are often conducting. Thus, one has to
chose the FM layers in such a way that it should be insulating in nature. To
show the multiferroism in materials, various physical characteristics can be
utilized, such as magnetocapacitance effect\cite{kimura,zheng,WP}.
Furthermore, properties of the superlattices basically depend on the various
structural details ({\it e.g.,} thickness, microstructure, composition,
strain,{\it \ etc.)} of the individual components and the interfaces among
them. Moreover, the lattice mismatch between the individual type of layer
materials as well as the one with the substrate, play important roles in
determining the interfacial structure, which in turn governs their
electronic and magnetic properties. Thus, it is necessary to study the
properties of superlattices composed of ferroelectric and ferromagnetic
components.

In previous works, we have studied superlattices composed of Pr$_{0.85}$Ca$%
_{0.15}$MnO$_3$/Ba$_{0.6}$Sr$_{0.4}$TiO$_3$ \cite{murga,murgaII,murgaIII}and
La$_{0.7}$Ca$_{0.3}$MnO$_3$/BaTiO$_3$ \cite{mpsingh} which behave as
multiferroics.\ In addition, it has been shown that their magnetoelectric
properties vary with the FE/FM layer thickness\cite
{murga,murgaII,murgaIII,mpsingh}. In this paper, we attempt to understand
the role of ferroelectric/paraelectric layers on multiferroism. To achieve
this goal, we have grown a series of superlattices, composed of the Ba$%
_{1-x} $Sr$_x$TiO$_3$ (BSTO), where {\it x} = 0, 0.2, 0.6, 0.8 and 1, and La$%
_{0.7}$Ca$_{0.3}$MnO$_3$ (LCMO), by utilizing the pulsed laser deposition
(PLD) process and characterized them by various techniques for their
structural and physical properties.

The growth of the superlattices were carried out at 720$^o$C in a flowing
100 mTorr O$_2$ ambient on (001)-oriented SrTiO$_3$ (STO)\cite{murga}. Based
on the previous work, showing the importance of the thickness of the layers
into the magnetoelectric properties of the LCMO-BTO superlattices\cite
{mpsingh}, we have limited our investigation to superlattices corresponding
to a total periodicity of 20 unit cells (u.c.). A peculiar attention was
paid for the LCMO layers, the latter were limited to the thickness of 5 u.c.
in order to get FM insulating behaviors\cite{Blamire} and BSTO layers was
fixed to 15 u.c. Finally, this structure was repeated 25 times in order to
get a total thickness of the order of 200 nm.

Crystallinity of the films was examined by X-ray diffraction (XRD).
Morphological study of the films was carried out by atomic force microscopy
(AFM). Magnetization (M) of the films was measured as a function of
temperature (T) and applied magnetic field (H) using a superconducting
quantum interference device magnetometer (SQUID). $DC$ electrical properties
of films were measured in four point probe configuration. To measure the
electrical properties of the films in current-perpendicular-to-the-plane
(CPP) geometry, a LaNiO$_3$ electrode was fabricated through a shadow mask 
\cite{murgaII}. Dielectric properties of the films were investigated by
employing a LCR meter and placing the samples in a PPMS system.

The detail XRD study shows the presence of satellite peaks along with the
main diffraction peaks leading to the chemical modulation of the order of
8.3 nm, which is very close to 15a$_P$ of BSTO + 5a$_P$ of LCMO (not shown),
where a$_P$ is the bulk lattice parameter of the individual perovskite
cells. The presence of higher order satellite peaks adjacent to the main
reflection peaks, arising from chemical modulation of multilayer structure,
indicates that the films were indeed grown coherently. The $\Phi $-scan
recorded around the $103$ reflection\cite{Phi} exhibits four peaks separated
by $90{{}^{\circ }}$ from each other (not shown), indicating a 4-fold
symmetry as expected for the perovskite structures of LCMO\ and BSTO,
confirming that the superlattices indeed grow epitaxially ''cube-on-cube''.
Fig 1a shows the {\it out-of-plane} lattice parameter of the various
superlattices, as extracted from the XRD data, as a function of Ba
concentration. For comparison, we have also plotted the lattice parameter of
bulk BSTO and the theoretical lattice parameter of the superlattices,
extracted from the formula [15a$_P$\ of (BSTO) + 5a$_P$\ of ( LCMO)]/20,
where a$_P$\ is the bulk lattice parameter of the individual perovskite
cells. From Fig. 1a, it is evident that with decreasing the Ba
concentration, the {\it out-of-plane} lattice parameter of superlattice
approaches its theoretical value. Fig.1b shows a typical 2D-AFM micrographs
of LCMO/BTO film and it reveals that the films are comprised of uniform
spherical size mounds having an average dimension of 65 nm. Furthermore,
morphological parameters, such as the root mean square roughness of the
superlattices, were extracted from the 2 x 2 $\mu $m$^2$AFM micrographs and
found to be of the order of 0.4 nm, {\it i.e.,} close to the perovskite unit
that indicates a very smooth surface.

The ($M-H$) curve exhibits a well-defined coercivity confirming the
ferromagnetic nature of the film.\ A typical magnetization {\it vs.}
magnetic field ($M-H$) loop for LCMO/Ba$_{0.6}$Sr$_{0.4}$TiO$_3$ film,
recorded at 10 K, is shown in Fig 2a. Furthermore, the
temperature-dependence of the magnetization $M(T)$ recorded under 3000\ Oe
applied magnetic field (inset of Fig 2a) shows the FM Curie temperature (FM-T%
$_C$) around 158\ K.\ This value is much lower than the observed bulk LCMO (%
\symbol{126}250 K) and it is mostly arising due to the substrate-induced
strain and the thickness of the layer (5\ u.c.)\cite{Blamire}. Thus, the
as-grown films are ferromagnetic in nature with FM-T$c$ in the range of 
\symbol{126}145-158 K. To study the coherent transport through these
structures and to examine the effect of the ferroelectric layers on
magnetoelectric properties of the superlattices, the resistance (R) was
measured under 0 {\it T} and 7 T magnetic field and the magnetoresistance ($%
MR$) is defined as $MR$ $(\%)=100\times [R(H)-R(0)]/R(0)$, where $R(H)$ and $%
R(0)$ are the resistance values measured with and without magnetic field,
respectively. It is well known fact that the intrinsic MR in manganites is
maximum near the FM-T$_C$ and therefore we have compared here the maximum MR
value observed in the various superlattices. Our study shows MR in the
superlattices strongly depends on the nature of the ferroelectric layer
employed for designing the multiferroic layers (Fig 2b). For example, the
superlattice composed of LCMO/BTO reveals a MR \symbol{126}$-30\%,$ whereas
a sharp decrease in the MR is observed by substituting a small amount of Sr
in the BTO layer,{\it \ e.g.,} the LCMO/Ba$_{0.6}$Sr$_{0.4}$TiO$_3$
superlattice exhibits MR \symbol{126}$-12\%$.

To understand the interaction between ferroelectricity and magnetism,
capacitance ($C$) measurements were performed versus temperature, at 1 KHz,
under 0 and 5 T. The $C(T)$ curves of these superlattices in 0{\it \ T} (Fig
3) exhibit a peak, suggesting ferroelectricity, whatever the Sr content up
to {\it x} = 0.40. This observation is in agreement with the fact that the
corresponding bulk titanates Ba$_{0.8}$Sr$_{0.2}$TiO$_3$ and Ba$_{0.6}$Sr$%
_{0.4}$TiO$_3$ were also found to be ferroelectric with the transition
temperature of 300 K and 250 K, respectively\cite{smol,book}. However, in
the present superlattices the transition temperatures are shifted towards
much lower values: one observes a transition temperature at \symbol{126}105
K for the LCMO/Ba$_{0.8}$Sr$_{0.2}$TiO$_3$ superlattice ( Fig 3a) and at 
\symbol{126}55 K for the LCMO/Ba$_{0.6}$Sr$_{0.4}$TiO$_3$ (Fig 3b). Hence,
the shift of the transition temperature with Sr content supports strongly
the presence of ferroelectricity in these films. This view point is
furthermore supported by the fact that the LCMO/SrTiO$_3$ superlattice (Fig
3c) does not show any peak of the capacitance in C(T) curve, in agreement
with the absence of ferroelectricity in SrTiO$_3$. The huge deviation of the
ferroelectric transition temperature in these superlattices with respect to
the bulk materials, can easily be understood by considering nature of
ferroelectric layers which are thin\cite{king} and consequently submitted to
large strains from the LCMO layers and indirectly also from the substrate%
\cite{karma}. One must indeed bear in mind that the ferroelectricity in the
perovskite structure depends on the nature of the distortion of the TiO$_6$
octahedra and the rattling of Ti$^{4+}$ inside the octahedra. As a
consequence, the ferroelectric transition temperature decrease drastically
as the size of the A-site cation decreases, e.g., tetragonal-BaTiO$_3$ is
ferroelectric, whereas cubic-SrTiO$_3$ is paraelectric, and CaTiO$_3$ which
exhibits another kind of distortion\cite{book}, is also paraelectric. Thus,
the La$^{2+}$/Ca$^{2+}$ cations located at the border of the BSTO perovskite
layers, may induce particular distortion in the BSTO layer, which represents 
\symbol{126}5\% of the $A$-sites of the ferroelectric layers, and may
decrease drastically the ferroelectric transition temperature.

The second important feature deals with the magnetocapacitance (MC), defined
as $MC$ $(\%)=100\times [C(H)-C(0)]/C(0)$, where $C(H)$ and $C(0)$ are the
capacitance measured with and without magnetic field, respectively. Fig
3(a)-(c) show the $C(T)$ curves for various superlattices measured at
different magnetic fields. One indeed observes that these superlattices
exhibit a negative $MC$ (Fig 3a-b). More importantly, the $MC$ value is
found to be maximum near the ferroelectric transition temperature, exactly
as the intrinsic magnetoresistance was found to be maximum near FM-T$_C$ in
the colossal-$MR$ manganites. In contrast, above the $FE$ transition
temperature the $MC$ disappears, as shown for instance at the FM-T$_C$ of
these superlattices (\symbol{126}150 K). These observations demonstrate that
the presence of ferroelectricity is absolutely necessary for the appearance
of magnetocapacitance. Importantly, $MC$ is very sensitive to the barium
content ( Fig 3d), {\it i.e.,} to the nature of the $FE$ layers, realizing
the highest value for the LCMO/BTO superlattice (i.e.{\it \ x} = 0),
decreasing abruptly for LCMO/Ba$_{0.8}$Sr$_{0.2}$TiO$_3$ ( {\it x} =0.2),
and so that no magnetocapacitance can be observed for the LCMO/SrTiO$_3$
superlattice ({\it x} =1), which does not exhibit any ferroelectricity. On
the other hand, the magnetic properties of the manganites are governed by
their structural distortion \cite{hwang}. In other words, by varying the Mn$%
^{3+}$-O-Mn$^{4+}$ bond angle, it is possible to tune the ferromagnetism in
manganites. The strain in the film plays a major role in determining the Mn$%
^{3+}$-O-Mn$^{4+}$ structure\cite{wil}. Hence, this study shows that in a
superlattice built up of ferromagnetic and ferroelectric layers,
magnetocapacitance can only be obtained in the temperature range where both
the ferromagnetism and ferroelectricity coexist. More importantly, for
designing a multiferroic lattice, it will be crucial to find an optimum
stress/strain in the film in order to induce the required interaction for
generating multiferroism. Thus, the presence of multiferroism in the
superlattices will be dependent on the details of FE and FM layers employed
for designing them.

To summarize, we have successfully grown LCMO/BSTO superlattices on
(001)-oriented SrTiO$_3$ by PLD. The ferroelectric transition temperature is
very much dependent on Ba composition. The physical measurements on these
superlattices show that the multiferroic properties of the superlattices
depend on the type of ferroelectric/paraelectric layers used. Finally, a
dependence of the ferroelectricity on the multiferroism has been clearly
demonstrated. Furthermore, this study provides a way to design new
multiferroics and understand their behaviors which will lead to the
realization of the devices based upon them.

We thank Dr.\ Ch.\ Simon for helpful discussions.\ This work has been
carried out in the frame of the Work Package ''New Architectures for Passive
Electronics'' of the European Network of Excellence ''Functionalized
Advanced Materials Engineering of Hybrids and Ceramics'' FAME (FP6-500159-1)
supported by the European Community, and by Centre National de la Recherche
Scientifique.

Figure 1: (a) : Out-of-plane experimental lattice parameter of the LCMO/BSTO
superlattices as a function of Ba content. The lattice parameters of bulk
BSTO and the theoretical values of the LCMO/BSTO superlattices are also
shown for comparison.\ (b) 2D: AFM micrograph of the LCMO/BaTiO$_3$
superlattices.

Figure 2: (a) ($M-H)$ loop of LCMO/Ba$_{0.6}$Sr$_{0.4}$TiO$_3$ superlattices
measured at 10 K.\ Inset shows the corresponding ($M-T)$ curve measured
under 3000 Oe applied magnetic field. (b) MR of the LCMO/BSTO superlattices
as a function of Ba composition.\ The solid line is just a visual guide.
Inset of Fig. 2b shows the $MR(H)$ of LCMO/BaTiO$_3$ superlattices measured
at 100K.

Figure 3. $C(T)$ curves of (a) LCMO/Ba$_{0.8}$Sr$_{0.2}$TiO$_3$, (b): Ba$%
_{0.6}$Sr$_{0.4}$TiO$_3$, and (c) LCMO/SrTiO$_3$ superlattices measured
under different applied magnetic fields. (d) Negative $MC$ of the LCMO/BSTO
superlattice as a function of Ba-composition.\ The solid line is a visual
guide to the eyes.


\begin{references}
\bibitem{Phi}  The indexation of the film is based on a perovskite subcell
assuming that the film is $c$-axis oriented.
\end{references}
\end{document}